\documentclass[
 aip, apl, amsmath, amssymb, reprint]{revtex4-1}
\usepackage{graphicx}
\usepackage{dcolumn}
\usepackage{amsmath}
\usepackage{bm}
\usepackage{subcaption}
\usepackage{multirow}
\usepackage[justification=raggedright]{caption}
\usepackage{float}
\usepackage[demo,abs]{overpic}
\usepackage{color,soul}
 \usepackage{float}
\makeatletter
\let\newfloat\newfloat@ltx
\makeatother
\usepackage{algorithm}
\usepackage{algpseudocode}
\usepackage{booktabs}
\usepackage{color,soul}
\usepackage{tabularx}
\usepackage{booktabs}
\usepackage{multirow}

\begin{document}

\title{Skull-Conforming Acoustic Holographic Lenses for Transcranial Targeting}

\author{Ceren Cengiz}
\author{Mihir Pewekar}
\author{Hrishikesh Kulkarni}
\affiliation{Department of Mechanical Engineering, Virginia Tech, Blacksburg, VA, $24061$, USA}
\author{Yunruo Ni}
\affiliation{Fralin Biomedical Research Institute, Virginia Tech, Roanoke, VA, $24016$, USA}
\author{Nathan Sambo}
\affiliation{Department of Mechanical Engineering, Virginia Tech, Blacksburg, VA, $24061$, USA}
\author{Adam Maxwell}
\affiliation{Department of Biomedical Engineering and Mechanics, Virginia Tech, Blacksburg, VA, $24061$, USA}
\author{Eli Vlaisavljevich}
\affiliation{Department of Biomedical Engineering and Mechanics, Virginia Tech, Blacksburg, VA, $24061$, USA}
\author{Wynn Legon}
\affiliation{Fralin Biomedical Research Institute, Virginia Tech, Roanoke, VA, $24016$, USA}
\author{Shima Shahab}
\email{sshahab@vt.edu}
\affiliation{Department of Mechanical Engineering, Virginia Tech, Blacksburg, VA, $24061$, USA}

\date{\today}

\begin{abstract}

Transcranial focused ultrasound (tFUS) offers noninvasive access to deep brain circuits but remains limited by skull-induced phase aberration, acoustic impedance mismatch, and poor volumetric control of intracranial pressure fields. Conventional phased-array and planar holographic strategies compensate aberrations electronically or computationally, yet do not resolve geometric and coupling inconsistencies imposed by subject-specific cranial morphology. We introduce personalized skull-conforming acoustic holograms that physically encode individualized wavefront corrections into a conformal acoustic interface. Within a subject-specific volumetric holography (SSVH) framework, cranial geometry and therapeutic constraints are embedded into a physics-based optimization pipeline for holographic phase synthesis. The resulting lens is integrated with a skull- and skin-conforming coupling layer that enhances impedance continuity, reduces reflection losses, and stabilizes spatial alignment, enabling simultaneous aberration mitigation and efficient transcranial transmission. Numerical simulations across multiple subjects and targets demonstrate consistent volumetric focusing and reliable target coverage while maintaining pressure fields within safety limits. Experimental validation using an ex vivo human skull confirms accurate fabrication, effective acoustic coupling, and faithful reconstruction of designed three-dimensional acoustic fields. By unifying wavefront engineering with anatomical conformity, this work establishes skull-conforming acoustic holography as a scalable strategy for high-fidelity, anatomically adaptive transcranial ultrasound targeting.

\end{abstract}

\keywords{Transcranial focused ultrasound, acoustic holography, acoustic hologram, holographic lens, neuromodulation, optimization}

\maketitle

\section{Introduction}

Transcranial focused ultrasound (tFUS) is a non-invasive technique that delivers focused acoustic energy to precisely targeted brain regions\cite{legon2024low, beisteiner2020transcranial}. Since its first application in humans in 2014 \cite{legon2014transcranial}, tFUS has attracted substantial research interest, particularly in the context of neurmodulation and therapeutic exploration  \cite{sarica2022human}. This growth has been especially notable in clinical applications targeting central nervous system (CNS) disorders, with a significant rise in published studies since 2020. These include investigations into disorders of consciousness \cite{yang2025promise, cain2022ultrasonic}, depression \cite{reznik2020double, fan2024thalamic}, epilepsy \cite{lee2022pilot, bubrick2024transcranial}, Parkinson’s disease \cite{samuel2023accelerated,osou2024novel}, chronic pain \cite{riis2024noninvasive, yu2020neuromodulation}, and others. Although these applications involve distinct therapeutic objectives, application-specific parameter choices, and varying safety considerations, they are unified by a common set of technical challenges rooted in acoustic wave propagation physics. These challenges include compensating for acoustic aberrations introduced by cranial and soft tissue structures \cite{Attali31122025}, establishing effective acoustic coupling between the ultrasonic source and the application region at the head–source interface \cite{tang2002investigation, lee2016transcranial, braun2020transcranial}, and achieving sufficient spatial coverage within the targeted region while suppressing off-target exposure \cite{fan2023computational, jin2024transcranial}. 

In current practice, single-element spherically focused transducers are commonly preferred for human tFUS applications due to their commercial availability, operational simplicity, and straightforward integration into existing clinical and research workflows. However, such systems inherently impose a predetermined focal geometry, limiting beam shaping flexibility and confining acoustic delivery to a fixed depth and volume \cite{legon2024low}. Under these constraints, skull-induced aberrations, subject-specific anatomical variability, and coupling-dependent effects remain largely unaddressed, limiting reliable control over the in situ pressure distribution \cite{freeman2025transcranial}. Subject-specific frameworks may offer a practical pathway toward overcoming these limitations by explicitly incorporating anatomical and acoustic variability into ultrasound delivery. The importance of such individualized strategies has been further underscored by the ITRUSST (International Consortium for Transcranial Ultrasonic Stimulation Safety and Standards) in their recent report outlining safety and efficacy considerations for clinical tFUS \cite{aubry2023itrusst}.

Advanced wavefront engineering strategies in tFUS include both electronically controlled phased-array systems and structurally encoded acoustic holography. While phased arrays offer dynamic beam steering, their inherent bulk, electronic complexity, and cost can limit practical deployment. In contrast, acoustic holography has emerged as a particularly powerful tool for subject-specific wavefield manipulation, enabling precise spatial control of acoustic energy delivery in tFUS applications. These include customized implementations such as acoustic metamaterials, phase plates, and 3D-printable lenses \cite{xie2016acoustic}. Among these, 3D-printable acoustic lenses, commonly referred to as acoustic holographic lenses (AHLs), have shown particular promise in therapeutic ultrasound applications \cite{jimenez2021acoustic, cengiz2024holographic, cengiz2025roadmap}. Acoustic holograms function by modulating the phase profile of an incoming wavefront, allowing reconstruction of a desired pressure distribution at the target region \cite{melde2016holograms, brown2020stackable}. In the context of tFUS, AHLs can be designed to encode subject-specific anatomical and acoustic features, enabling design optimization based on individual variability \cite{maimbourg2019steering, sallam2024gradient, jimenez2019holograms}.

Despite these promising capabilities, the practical deployment of acoustic holography in tFUS remains constrained by computational design complexity, limited volumetric targeting flexibility, and the absence of application-driven optimization frameworks that account for subject-specific anatomical and therapeutic constraints. Addressing these gaps, this work introduces a subject-specific volumetric holography (SSVH) framework aimed at translating acoustic holography-based ultrasound delivery toward robust preclinical and clinical deployment. The objective is to compute a unique phase plate that modulates the wavefront emitted from a planar pressure source to generate a targeted volumetric pressure field at a prescribed location after transmission through the acoustically heterogeneous layers of the head, including the skin, skull, and brain. The proposed framework builds upon the fundamental principles of computer-generated holography (CGH) \cite{pi2022review, shi2021towards} and integrates optimization-based strategies to enhance wavefield control \cite{zhang20173d, melde2023compact}.

Prior work from our group introduced a gradient-descent-based optimization approach combined with a modified mixed-domain method (MMDM) for tFUS holography \cite{sallam2024gradient}, demonstrating the feasibility of optimization-driven phase design under controlled conditions using a simplified, 3D-printed skull model with minimal aberration effects. Building on this foundation, the proposed SSVH framework advances acoustic holography beyond proof-of-concept configurations by explicitly incorporating subject-specific anatomical variations, application-driven constraints, and target depth variability into the optimization pipeline while providing a pathway toward practical device configuration and manufacturing. The framework is formulated to support individualized optimization across different subjects and targeting scenarios, accounts for realistic skull-induced heterogeneity, and is experimentally validated using ex vivo human skull specimens rather than idealized 3D-printed skull phantoms. Critically, a persistent gap in existing acoustic holography studies for tFUS is the assumption of idealized water-coupled conditions, which renders proposed designs incompatible with real-world application environments. To bridge this gap, the SSVH framework is accompanied by a conformal, skull- and skin-adaptive coupling layer and an associated manufacturing strategy, establishing a complete, deployment-ready package for holography-assisted tFUS with direct translational impact across research and clinical environments.

\section{Methodology}

\subsection{Subject-specific domain formation}

Since human head anatomy varies significantly across individuals \cite{lillie2016evaluation, alexander2019structural, riis2022acoustic}, subject-specific solutions were achieved by processing anatomical data and converting it into a 3D matrix form suitable for computational modeling. For this, patients’ computed tomography (CT) scans were used to extract skull geometry and magnetic resonance imaging (MRI) scans were used to define the target regions. MRI data were acquired on a Siemens 3T Prisma scanner (Siemens medical Solutions, Erlangen, Germany) at the Fralin Biomedical Research Institute's Human Neuroimaging Laboratory. Structural images were collected using a T1-weigted 3D MPRAGE sequence with the following parameters: repetition time (TR) = 2.30 s, echo time (TE) = 2.32 ms, inversion time (TI) = 900 ms, flip angle = 8°, slice thickness = 0.9 mm, isotropic voxel size = 0.9 × 0.9 × 0.9 mm, matrix size = 256 × 256, phase field of view = 100 \%, parallel imaging (GRAPPA) factor = 2, and sagittal acquisition orientation. CT scans were collected with a Kernel = Hr60 in the bone window, FoV= 250 mm, kilovolts (kV) = 120, rotation time = 1s, delay = 2S, pitch = 0.55, caudocranial image acquisition order, 1.0 mm iamge increments for a total of 121 images and scan time of 13.14s.

Brain segmentation and region-of-interest  (ROI) definition were performed using a combination of subject-specific and atlas-based approaches, depending on anatomical region. Hippocampal segmentation was obtained from each participant's T1-weighted anatomical image using FreeSurfer (recon-all; v7.4.1) \cite{fischl2012freesurfer}, yielding automated subject-specfic hippocampal labels in native T1 space. Other ROIs were defined using established probabilistic atlases in Montreal Neurological Insitutue (MNI) space. Specifically, cuneus was defined using the Harvard-Oxford cortical structural atlas \cite{desikan2006automated}, and the insualar cortex was defined using the functinoal parcellation described by Deen et al.\cite{deen2011three}. These atlas-based ROIs were nonlinearly transformed from MNI space into each participant's native anatomical space using Advanced Normalization Tools (ANTs)\cite{avants2011reproducible}. Following ROI segmentation and co-registration, candidate source pathways were identified based on the shortest distance to each target while respecting anatomical constraints. After determining the source location, a skull-conforming layer was incorporated into the domain and modeled according to the subject-specific skull morphology and overlying skin structure. The skin, skull, brain, conforming layer, and target were defined using numerical identifier to facilitate computational analysis.

\renewcommand{\arraystretch}{1.2}
\setlength{\tabcolsep}{4pt}
\begin{table}
\centering
\caption{Subject demographics and target specifications. $V_{t}$ = target volume (cm$^{3}$); $D_x$, $D_y$= lateral distances from the geometric center of the transducer (cm); $D_z$ = axial distance from the transducer surface (cm). $h_{skull}$ = mean skull thickness along the acoustic propagation direction (cm). Group values are reported as mean $\pm$ SD.}
\label{tab:targets_topdown}
\begin{tabular}{lcccccc}
\hline
 & \multicolumn{5}{c}{\textbf{Subject \#}} & \textbf{Group} \\
\cline{2-6}
 & 1 & 2 & 3 & 4 & 5 & \\
\hline

\textbf{Age (year)} &20&21&19&55&50&33 $\pm{17.9}$\\

\textbf{Sex} & M & M & M & F & M & \\
\hline
\multicolumn{7}{l}{\textbf{Deep target: Hippocampus (Hippo)}} \\
\quad $V_{t}$        & 8.3 & 7.7 & 8.0 & 7.8 &8.3 & 8 $\pm{0.3}$  \\
\quad $D_x$      & 0.1 & 0.1 & 0.4 & 0.4 & 0.1 & 0.2 $\pm{0.2}$  \\
\quad $D_y$      & 0.1 & 2.1 & 2.6 & 2.6 &2.6 & 2 $\pm{1.1}$  \\
\quad $D_z$      & 8.9 & 8.8 & 8.5 & 8.5 & 8.6 &  8.7 $\pm{0.2}$ \\
\quad $h_{skull}$  &0.8 &0.7 & 0.4 & 0.6 &0.8 & 0.7 $\pm{0.2}$
\\
\hline
\multicolumn{7}{l}{\textbf{Mid-depth target: Left anterior insula (LAI)}} \\
\quad $V_{t}$         & 4.2 & 4.3 & 4.7 & 3.3 & 3.6 & 4 $\pm{0.6}$ \\
\quad $D_x$      & 0.1 & 0.04 & 0.1 & 0.1 & 0.1 & 0.1 $\pm{0.02}$ \\
\quad $D_y$      & 0.1 & 0.04 & 0.1 & 0.1 & 0.1 & 0.1 $\pm{0.02}$ \\
\quad $D_z$      & 7.2 & 7.5 & 7.4 & 7.2 & 7.3 & 7.3 $\pm{0.1}$  \\
\quad $h_{skull}$ & 0.5 &0.5 &0.5 &0.6 & 0.5 & 0.5 $\pm{0.05}$
\\
\hline
\multicolumn{7}{l}{\textbf{Superficial target: Cuneus}} \\
\quad $V_{t}$         & 20.7 & 17.6 & 16.8 & 10.8 & 15.5 & 16.3 $\pm{3.9}$ \\
\quad $D_x$      & 0.1 & 0.1 & 0.1 & 0.1 & 0.1 & 0.1 $\pm{0}$ \\
\quad $D_y$      & 0.1 & 0.1 & 0.1 & 0.1 & 0.1 &0.1 $\pm{0}$  \\
\quad $D_z$      & 6.4 & 6.2 & 5.8 & 6.9 & 4.7 & 6 $\pm{0.79}$ \\
\quad $h_{skull}$ &1.0 & 0.8 & 0.6 & 1.1 &0.8 & 0.9 $\pm{0.2}$ 
\\
\hline
\multicolumn{7}{l}{\textbf{Superficial target: Left cuneus (Cuneus-L)}} \\
\quad $V_{t}$         & 10.5 & 8.6 & 9.9 & 5.7 & 8.5 & 8.6$\pm{1.8}$ \\
\quad $D_x$      & 0.1 & 0.1 & 0.1 & 0.1 & 0.1 &  0.1 $\pm{0}$ \\
\quad $D_y$      & 0.1 & 0.1 & 0.1 & 0.1 & 0.1 & 0.1 $\pm{0}$ \\
\quad $D_z$      & 5.4 & 6.2 & 5.7 & 5.7 & 5.4 & 5.7 $\pm{0.3}$\\
\quad $h_{skull}$ &1.0 &0.8 &0.6 &1.1 &0.8 & 0.9 $\pm{0.2}$ 
\\
\hline
\end{tabular}

\end{table}

\begin{figure*}
    \centering
    \includegraphics[width=0.8\linewidth]{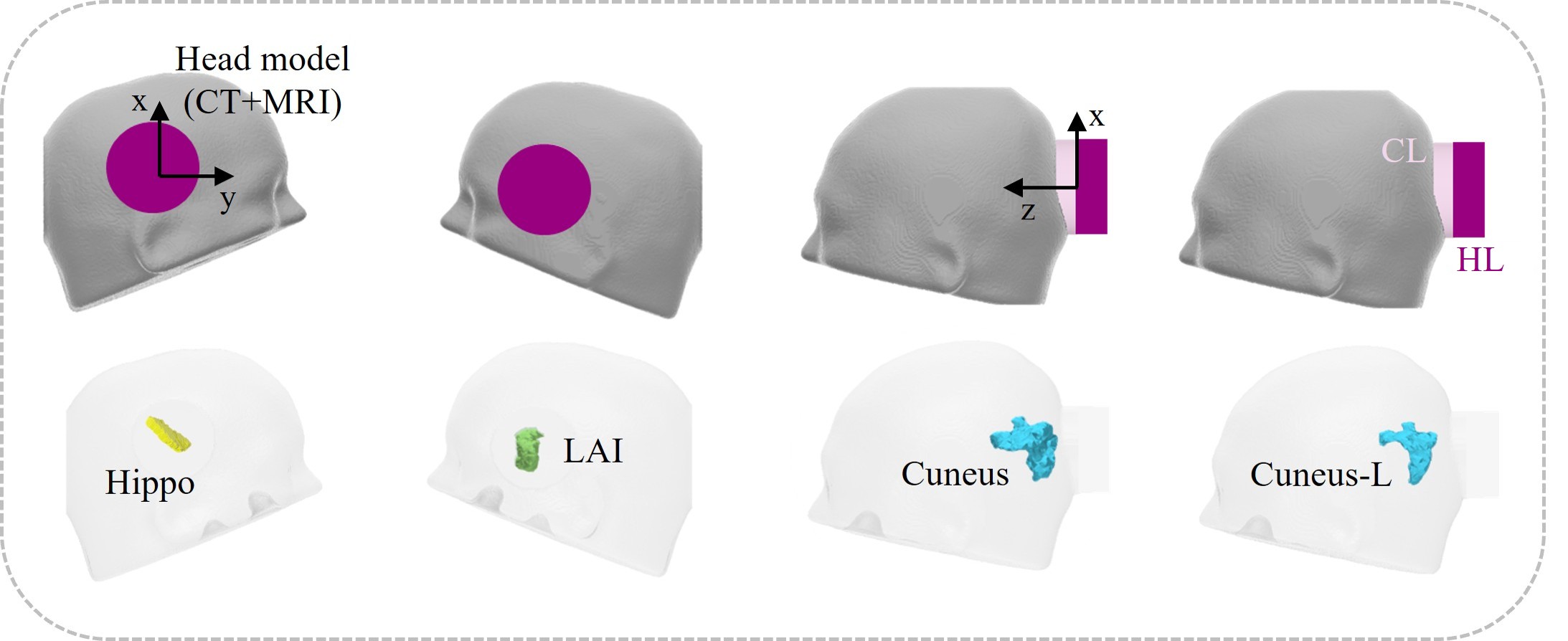}
    \caption{CT-MRI co-registered head model of Subject \#5 illustrating four representative target brain regions (Hippocampus (Hippo), Left anterior insula (LAI), Cuneus, Left cuneus (Cuneus-L)) and their corresponding source pathways. CL: Conforming layer and HL: Hologram layer.}
    \label{fig:subject5targets}
\end{figure*}

Table \ref{tab:targets_topdown} summarizes demographics and target speficiations considered in this study. Five subjects were considered to capture subject-specific variations in anatomy and propagation conditions, with Subjects 1–3 within 18-45 age range and Subjects 4-5 above 45 years of age. To comprehensively evaluate the SSVH system's performance across varying brain depths and potential clinical applications of tFUS, four targets located within three representative brain regions were investigated. 
The hippocampus serves as a deep target (associated with memory and learning) \cite{eichenbaum2004hippocampus}, the insula as a mid-depth target (involved in sensory and emotional processing) \cite{gogolla2017insular}, and the cuneus as a superficial target (commonly investigated in visual processing studies) \cite{vanni2001coinciding}. Considering the anatomical extent of the cuneus, both the full cuneus and its left subregion are examined to capture potential variations in sonication performance across the cortical area. Target-subject configurations are defined by the target volume ($V_{t}$), lateral source-target offsets ($D_x$, $D_y$), axial distance from the transducer surface to the center of gravity of the target ($D_z$), and the mean skull thickness along the acoustic propagation direction ($h_{skull}$). A representative visualization of the target locations and corresponding source pathways for one subject is shown in Fig. \ref{fig:subject5targets}.

\subsection{Hierarchical hologram optimization (HHO)}

Once the subject-specific computational domain is established, the next step is to solve the wave propagation in the heterogeneous medium and compute the thickness map for the 3D-printable AHL design. This tailored thickness profile modulates the transmitted wavefront to reconstruct the desired complex acoustic field with a targeted spatial pressure distribution. When the model accounts for skull-induced aberrations and enforces a volumetric target profile, the complex thickness profile can encode all the necessary information to achieve the desired tFUS outcome. In this study, MMDM implemented in the open-source MATLAB toolbox mSOUND \cite{gu2021msound} was employed to enable computationally efficient forward and backward wave propagation in a heterogeneous domain. MMDM extends the mixed-domain method (MDM) \cite{gu2018numerical} by introducing phase and amplitude corrections. It accounts for wavefront transmission and reflection effects, and has been demonstrated to accurately model wave propagation through skull-like aberrating layers in heterogeneous media \cite{gu2020modified}. Determining the optimal thickness map requires multiple forward and backward wave propagations while incorporating source boundary conditions and desired target region coverage. This process can be facilitated by optimization methods that iteratively adjust the phase map to satisfy the enforced error and loss function criteria. The gradient-based optimization strategy employed in this study builds upon our prior work, which established the feasibility of optimization-driven phase design within the MMDM framework for tFUS holography \cite{sallam2024gradient}. Here, we extend this approach by implementing a two-stage coarse-to-fine optimization strategy, referred to as hierarchical hologram optimization (HHO). This formulation enables efficient wavefield modeling in strongly heterogeneous media while systematically incorporating subject-specific anatomical information and application-driven constraints into the computational pipeline.

\begin{figure*}
    \centering
    \includegraphics[width=1\linewidth]{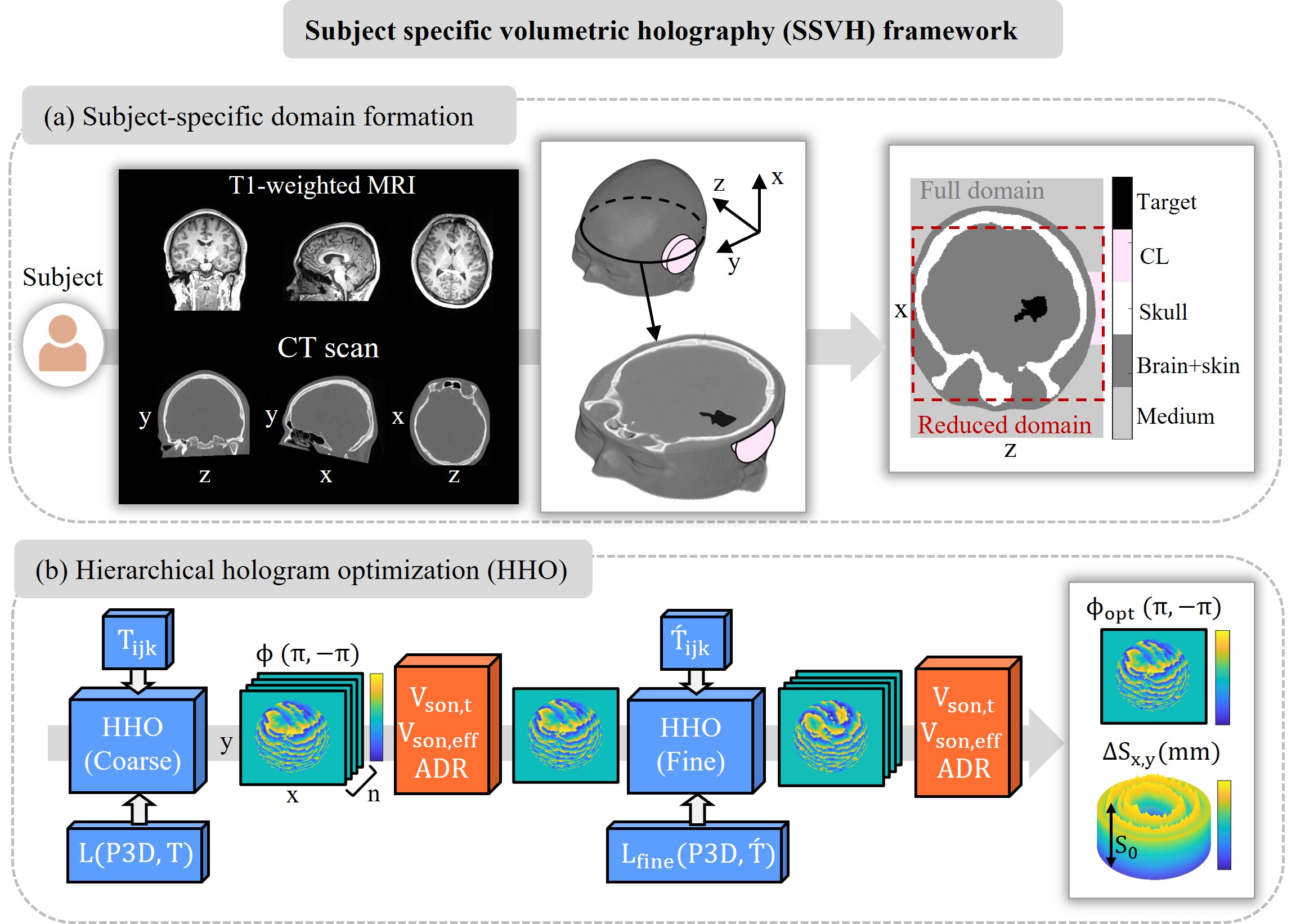}
    \caption{Methodology of the subject-specific volumetric holography (SSVH) framework. 
(a) Subject-specific domain formation: MRI/CT acquisition; brain and skull segmentation; MRI–CT coregistration; computational domain construction (CL: Conforming layer). 
(b) Hierarchical hologram optimization (HHO): Phase map ($\phi$) optimization under subject-specific constraints and targeting requirements; efficiency metrics for performance evaluation ($V_{\text{son,t}}$: sonicated target volume, $V_{\text{tvf}}$: targeted volume fraction, $ADR$: amplification demand ratio); thickness map ($\Delta S_{xy}$) calculation.}
    \label{fig:methodology}
\end{figure*}

In the coarse optimization stage, the phase distribution $\phi$ at the hologram plane was obtained by minimizing the discrepancy between the generated volumetric acoustic field $P3D$ and the prescribed target volume $T$, while enforcing the a fixed pressure amplitude at the transducer plane $|P_0|=A_0$. The squared loss function $L$ and its gradient $\nabla L$ used in this stage are defined as follows \cite{melde2023compact}:

 \begin{equation}
L(P3D, T)=\frac{1}{2} \sum_{i=1}^{n_{x}} \sum_{j=1}^{n_{y}} \sum_{k=1}^{n_{z}}\left(P3D_{i j k}-T_{i j k}\right)^{2}
\end{equation}

\begin{equation}
\nabla L=2 A^{*}\left(P3D-T \circ \frac{P3D}{|P3D|}\right)
\end{equation}

Where $A^*$ denotes the backward propagation from each plane in the volume back to the transducer plane such that $A^{*}P3D=P_0$.

In the fine optimization stage, the coarse solution was further refined through two complementary mechanisms: target refinement and adaptive loss formulation. Target refinement involves modifying the prescribed target volume $T$ to improve optimization tractability and targeting performance, for example by applying controlled geometric smoothing to mitigate uneven or fragmented structures that can hinder convergence or lead to undesired sonication patterns. In parallel, customized loss formulations were employed by augmenting the baseline objective with auxiliary penalty terms designed to regulate acoustic energy outside the intended target region. These penalties are informed by monitoring optimization behavior over the target volume and selected orthogonal central planes, together with inspection of the corresponding pressure amplitude distributions to assess the need for direction-dependent suppression. In this manner, subject-specific anatomical features and target characteristics were systematically incorporated into the fine-stage HHO optimization process.

The squared loss function for the fine optimization stage can be expressed as:

\begin{align}
L_{fine}(P3D, \overset{\prime}{T})
&= C_1 \cdot \tfrac{1}{2} \sum_{i=1}^{n_x}\sum_{j=1}^{n_y}\sum_{k=1}^{n_z} 
   \left(P3D_{ijk}-\overset{\prime}{T}_{ijk}\right)^2 \nonumber \\
&\quad + L_{1} + L_{2} + L_{3}
\end{align}

Where $C_{1}$ is the weight assigned to the loss function in the whole domain, and $\overset{\prime}{T}$ denotes the refined target volume. The terms $L_{1}$, $L_{2}$ and $L_{3}$ represents the additional penalty terms incorporated into the fine loss function along the three orthogonal planes. The calculation of these penalty terms is given by:

\begin{equation}
L_{1}= C_{2}\cdot \frac{1}{2}\sum_{i=1}^{n_x}\sum_{j=1}^{n_y}(P3D_{ijk_0}-\overset{\prime}{T}_{ijk_0})^2
\end{equation}

\begin{equation}
L_{2}= C_{3}\cdot \frac{1}{2}\sum_{i=1}^{n_x}\sum_{k=1}^{n_z}(P3D_{ij_0k}-\overset{\prime}{T}_{ij_0k})^2
\end{equation}

\begin{equation}
L_{3}= C_{4}\cdot \frac{1}{2}\sum_{j=1}^{n_x}\sum_{k=1}^{n_z}(P3D_{i_0jk}-\overset{\prime}{T}_{i_0jk})^2
\end{equation}

Where $C_{2}$, $C_{3}$ and $C_{4}$ denote the weights assigned to each orthogonal plane. In this work, additional penalization is applied along the three principal orthogonal planes for simplicity; however, the loss formulation can be readily extended to include alternative planes when required by the application.

The phase distribution $\phi$ obtained from the coarse optimization stage serves as the initial condition for the fine optimization. This warm-start strategy leverages the best-performing phase map from the preceding stage to improve convergence and stability. Specifically, the initialization $\phi$ was selected by evaluating tFUS targeting effectiveness across the coarse-stage iterations and retaining the phase distribution that yields the highest performance. Following completion of the fine optimization stage, targeting effectiveness was re-evaluated to identify the final optimal phase distribution $\phi$, which is subsequently used to generate the AHL thickness map.

Normalized pressure amplitudes were employed throughout the optimization stages of the SSVH framework to promote numerical stability and facilitate convergence. Targeting effectiveness was subsequently assessed by computing the sonicated volumes inside and outside the prescribed target region. This evaluation requires rescaling the normalized pressure fields to physically meaningful pressure amplitudes relevant for effective tFUS neuromodulation. Based on prior study demonstrating successful dorsal anterior cingulate cortex (dACC) neuromodulation at peak pressure amplitudes of $115.2\pm{53.1}$ kPa \cite{strohman2024low}, a reference neuromodulation pressure amplitude of 115 kPa is adopted in this study.

To translate the normalized pressure fields into physically meaningful pressure amplitudes, a representative pressure amplitude within the target region was identified using the statistical mode of the pressure distribution. Specifically, pressure amplitudes within the target mask are extracted and binned into a histogram, and the center of the bin with the highest occurrence was selected as the dominant pressure value. This value was then scaled to correspond to a reference pressure amplitude of 115 kPa. Following rescaling, the maximum pressure amplitude within the brain volume was recorded for safety evaluations. Although no FDA-approved guidelines currently exist that define ultrasound exposure limits specific to tFUS neuromodulation, recently published recommendations from the ITRUSST initiative provide an explanatory framework for safety in tFUS, including contraints based on the mechanical index ($MI$ $<$ 1.9; a dimensionless parameter estimating the likelihood of cavitation-related bioeffects) \cite{aubry2025itrusst}. 

To ensure consistent acoustic exposure across subjects, targeting effectiveness in this study was evaluated by delivering the same dominant pressure amplitude to each case. The dominant pressure value is used to define a threshold, $P_{th}$, and regions in which the pressure amplitude is equal to or greated than $P_{th}$ were classified as successfully sonicated and used to quantify targeting performance. 

The sonicated volume within the target region was calculated as:

\begin{equation}
V_{\text{son,t}}= \sum_{i=1}^{n_x}\sum_{j=1}^{n_y}\sum_{k=1}^{n_z} (P3D_{ijk}>P_{th}\cdot T_{ijk}=1)\cdot v
\end{equation}

where $T_{ijk}$ is a binary mask defining the region that encompasses the target, $v$ is the unit voxel volume, computed as $v = dx \cdot dy \cdot dz$. $V_{\text{son,t}}$ in cubic centimeters ($cm^3$), while \(V_{\text{son,t}}(\%)=S_{\text{son,t}}/V_{t}\times100\) represents the percentage of the target region that is successfully sonicated, where $V_{t}$ denotes the total volume of the target. The total sonicated volume within the brain, including both inside and outside the target domain, was calculated as:

\begin{equation}
 V_{\text{son,b}}= {\sum_{i=1}^{n_x}\sum_{j=1}^{n_y}\sum_{k=1}^{n_z}(P3D_{ijk} > P_{\text{th}} \cdot E_{ijk} = 1) \cdot v}  
\end{equation}

where $E_{ijk}$ is a binary mask defining the region that encompasses the target and surrounding areas within the brain domain. The targeted volume fraction, $V_{\text{tvf}}$ quantifies the total acoustic field concentrated within the target region.

\begin{equation}
 V_{\text{tvf}}(\%)=\frac{ V_{\text{son,t}}}{ V_{\text{son,b}}}\times100
\end{equation}

The amplification demand ratio ($ADR$) quantifies the relative power requirement of a given AHL design and was defined as the ratio between a reference pressure amplitude of 115 kPa and the dominant normalized pressure amplitude used as the threshold. A lower $ADR$ indicates that less amplification is required to achieve the reference pressure level, corresponding to reduced power demand from the ultrasound source in tFUS operation. Because all evaluations were performed using normalized pressure amplitudes, $ADR$ functions as a relative metric that captures the amplification characteristics of the system and was used to compare and identify AHL designs with more favorable power requirements. 

\begin{equation}
    ADR= \frac{115}{P_{th}}
\end{equation}

After obtaining a satisfactory hologram phase distribution $\phi$, it was converted into a thickness map matrix that specifies the spatial thickness profile required to reconstruct the desired acoustic field (see Eq.\ref{THICKNESSMAP} \cite{melde2016holograms}). This thickness map was subsequently exported in Standard Tessellation Language (STL) format and used for 3D printing of the AHL.

\begin{equation}
    \label{THICKNESSMAP}
    \Delta S_{x,y}=S_{0}-\frac{\Delta \phi_{x,y}}{(k_m-k_h)}
\end{equation}

Where the pixel thickness is denoted by $ \Delta S_{x,y}$, the phase difference by $\Delta \phi_{x,y}$, the initial thickness of the hologram plane by $S_0$, and the wavenumbers of the hologram and surrounding medium by $k_h$ and $k_m$ respectively.

Since the analysis were carried out for different patients with subject-specific computational domains, the simulations were run on grids of varying sizes ($n_x$, $n_y$, $n_z$).To enhance computational efficiency, the domain size was reduced when feasible, with the largest grid size being $256 \times 320\times 512$ . A spatial resolution of $d_x=d_y=0.46$ mm and $d_z=0.45$ mm, corresponding to approximately $\lambda_w/4.8$ in water and $\lambda_s/9.2$ in skull was found to be sufficient for the analysis, and second-order reflections were used. Perfectly matched layers (PML) were added to the boundaries of the computational domain to suppress unwanted reflections. Skull acoustic properties were assumed to be constant, with values defined as \(c_{skull} = 2850 \)  m/s, \(\rho_{skull} = 1732\)  kg/m$^3$, \(\alpha_{skull}  = 7.38 \) dB/cm.MHz \cite{mueller2017numerical}.

\begin{figure*}
    \centering
    \includegraphics[width=1\linewidth]{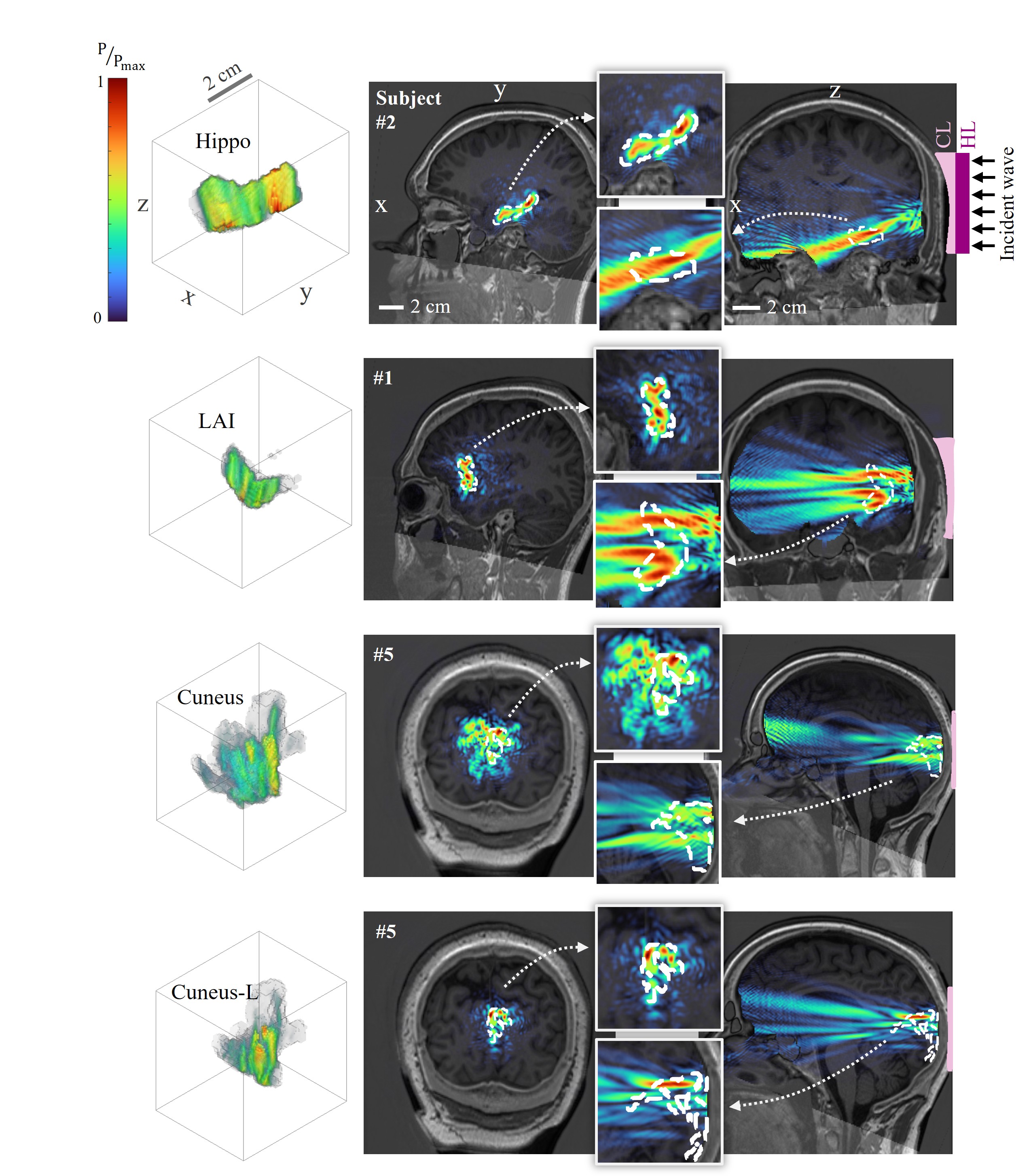}
    \caption{Targeting results for five brain regions. The rows correspond, from top to bottom, to the deep target Hippocampus (Hippo) for Subject \#2, the mid-depth target left anterior insula (LAI) for Subject \#1, the superficial target Cuneus for Subject \#5, and the superficial target left Cuneus (Cuneus-L) for Subject \#5.}
    \label{fig:compres}
\end{figure*}

\section{Computational Results}

 Four distinct brain regions were investigated in this study (see Table \ref{tab:targets_topdown}). The resulting values of $V_{\text{son,t}}$, $V_{\text{tvf}}$, and $ADR$ are summarized in Table \ref{finalresults}. Overall, the SSVH framework yields comparable targeting performance across cases, with mean values of $V_{\text{tvf}}=23\%\pm{2.5}$, $V_{\text{son,t}}=45\%\pm{9.6}$, and $ADR=410\pm{91.4}$. 
 
 Across the four brain locations considered, the highest $V_{\text{tvf}}$ values were obtained for the Cuneus-L, closely followed by Hippo. Inter-subject variability differs between these cases, with Hippo exhibiting more consistent $V_{\text{tvf}}$ values across subjects. The highest $V_{\text{son,t}}$ values were observed for superficial locations, consistent with their proximity to the skull, such that partially defocused regions of elevated pressure remain within the prescribed target volume. Correspondingly, the lowest $ADR$ values, indicating reduced amplification demand, were obtained for Hippo, followed by LAI. In contrast, superficial cases exhibit increased wavefield spreading and greater skull interaction, which likely contribute to the higher amplification demand observed for these configurations.

\begin{table*}
\centering
\caption{Summary of the targeting results at the end of the HHO process. $V_{\text{son,t}}$ $(cm^3)$ $=$ sonicated volume $(>115 kPa)$, $V_{\text{son,t}}$ $(\%)$ $=$ $\%$ of the target volume successfully sonicated, $V_{\text{tvf}}
$ $(\%)$ = targeted volume fraction, $ADR$ $=$ amplification demand ratio. Group values are reported as mean $\pm$ SD.}
\renewcommand{\arraystretch}{1.2}
\setlength{\tabcolsep}{6pt}
\label{finalresults}
\begin{tabular}{llcccccc}
\hline
& \multicolumn{6}{c}{\textbf{Subject \#}} & \textbf{Group} \\
\cline{3-7}
&  & 1 &2& 3&4 & 5 \\
\hline
\multirow{4}{*}{\textbf{Deep target: Hippo}} 
& $V_{\text{son,t}}$ $(cm^3)$ & 4.2 & 2.0 & 2.0 & 2.1 & 2.4 & $2.5 \pm{0.9}$ \\
& $V_{\text{son,t}}$ $(\%)$ & 50.3 & 26.4 & 25.5 & 26.4 & 28.9 & $31.5 \pm{10.6}$ \\
& $V_{\text{tvf}}$ $(\%)$ & 25.2 & 28.7 & 21.8 & 25.3 & 22.6 & $24.7 \pm{2.7}$ \\
& $ADR$ & 317.2 & 249.9 & 284.1 & 262.5 & 379.0 & $298.5 \pm{51.7}$ \\

\hline
\multirow{5}{*}{\textbf{Mid-depth target: LAI}} 
& $V_{\text{son,t}}$ $(cm^3)$ & 1.2 & 2.0 & 2.8 & 1.4 & 1.9 & $1.8 \pm{0.6}$ \\
& $V_{\text{son,t}}$ $(\%)$ & 27.5 & 57.2 & 58.4 & 41.5 & 43.7 & $45.7 \pm{12.7}$ \\
& $V_{\text{tvf}}$ $(\%)$ & 31.0 & 21.4 & 15.5 & 21.0 & 23.7 & $22.5 \pm{5.6}$ \\
& $ADR$ & 253.8 & 429.3 & 628.9 & 276.0 & 376.6 & $392.9 \pm{150.3}$ \\

\hline
\multirow{5}{*}{\textbf{Superficial target : Cuneus}} 
& $V_{\text{son,t}}$ $(cm^3)$ & 7.9 & 4.7 & 10.1 & 5.1 & 8.2 & $7.2 \pm{2.3}$ \\
& $V_{\text{son,t}}$ $(\%)$ & 38.4 & 54.9 & 60.5 & 46.8 & 53.1 & $50.7 \pm{8.5}$ \\
& $V_{\text{tvf}}$ $(\%)$ & 21.6 & 21.1 & 15.8 & 15.1 & 23.7 & $19.5 \pm{3.8}$ \\
& $ADR$ & 504.0 & 407.6 & 631.7 & 556.8 & 496.2 & $519.3 \pm{82.6}$ \\
\hline
\multirow{5}{*}{\textbf{Superficial target: Cuneus-L}} 
& $V_{\text{son,t}}$ $(cm^3)$ & 5.4 & 4.7 & 4.6 & 3.5 & 4.2 & $4.5 \pm{0.7}$ \\
& $V_{\text{son,t}}$ $(\%)$ & 51.9 & 54.9 & 46.5 & 61.4 & 49.8 & $52.9 \pm{5.6}$ \\
& $V_{\text{tvf}}$ $(\%)$ & 25.2 & 27.8 & 20.3 & 14.9 & 37.0 & $25.1 \pm{8.3}$\\
& $ADR$ & 437.7 & 407.6 & 410.0 & 523.7 & 373.6 & $430.5 \pm{56.8}$ \\
\hline
\end{tabular}
\end{table*}

Representative pressure amplitude distributions corresponding to the cases with the highest $V_{\text{tvf}}$ values for each brain region are shown in Fig. \ref{fig:compres}. In the xy plane, the pressure fields exhibit confinement within the prescribed target regions, indicating effective focal coverage. In the xz plane, off-target sonication is more apparent, reflecting the three-dimensional spreading of the acoustic field outside the target volume. Notably, for Hippo Subject $\#2$, misalignment of the acoustic pathway is visible as a consequence of ear-region avoidance during application (see Table~\ref{tab:targets_topdown}). Although localized off-target sonication is present, the pressure distribution exhibits effective beam shaping toward the intended target. This observation underscores the capacity of the proposed SSVH optimization to operate under constrained application pathways while preserving focal coverage, thereby offering increased flexibility relative to conventional tFUS system designs.

While Table \ref{finalresults} reports the final targeting performance achieved with the SSVH framework, comparison of the targeted volume fraction obtained after the HHO coarse and fine stages elucidates the contribution of fine optimization. Across all analyzed cases, the fine optimization stage results in an increase in $V_{\text{tvf}}$ of $19.4 \% \pm{5.5}$ for Hippo, $10.6 \% \pm{16.2}$ for LAI, $2.3 \% \pm{4.1}$ for Cuneus and $8.8 \% \pm{18.5}$ for Cuneus-L. The magnitude of improvement varies across targets, with comparatively smaller improvements observed for superficial regions, which is consistent with their geometric complexity and extended spatial extent relative to deeper targets.

Overall, incorporation of subject-specific HHO fine optimization within the SSVH framework enables the determination of AHL designs that achieve reliable and comparable targeting performance for tFUS applications across diverse anatomical conditions. At the same time, variations in targeting outcomes persist across subjects and brain regions. These differences reflect the combined influence of subject-specific anatomical factors, including skull thickness, local skull geometry, target volume and spatial extent, anatomical location, and regional brain morphology, which collectively govern the achievable targeting performance.

\section{Fabrication and Experimental Methodology}

All experiments were conducted in a submerged water-tank configuration to provide stable acoustic medium and enable quantitative mapping of the transmitted pressure field through ex-vivo human skull bone. 

\subsection{CT-based personalization and subject-specific holographic lens fabrication.}
For the skull specimen and targeting condition, a subject-specific acoustic holographic lens was designed to compensate skull-induced phase distortions at the operating frequency. Skull morphology was obtained from Hounsfield units based thresholding of the CT scan (scanner/model: SOMATOM Confidence scanner (Siemens Healthineers, Erlangen, Germany); voxel size: $0.386\,\mathrm{mm}$), shown volumetrically in Fig.~\ref{fig:experimental}(a.1). The CT-derived skull morphology was used to define (i) the desired target location and/or target plane (i) the shape of the conformal layer that adapts to the complex morphology of the skull, and (iii) the formation of the computational domain as an input to the HHO algorithm. 

\begin{figure*}[htp]
    \centering
    \includegraphics[width=1\linewidth]{fab.png}
    \caption{Schematic depicting the fabrication process of a subject-specific holographic lens. 
(a) SLA-printed acoustic hologram fabricated from the thickness map obtained using the HHO method. 
(b) Casting process in which biocompatible resin is sequentially injected to form the lens structure. 
(c) Illustration of the stepwise curing process using 405~nm UV light to solidify each resin layer. 
(d) SLA-printed subject-specific skull conformal layer designed to match the outer skull surface. 
(e) Fully fabricated lens assembly showing the integrated hologram and conformal layer. 
(f) Exploded-view CAD model illustrating the assembly components, including the transducer, holder, hologram, and conformal layer. 
(g) Final assembled CAD model of the complete subject-specific holographic lens system.}
    \label{fig:Fabrication}
\end{figure*}

The lens system comprises a skull-conformal layer and a hologram, with additional supporting features to enable application in a transcranial neuromodulation setting. Formlabs Clear Resin (V4) was used for the hologram, while the biocompatible BioMed Elastic 50A resin was used for the skull-conformal layer. The lens was fabricated using a Formlabs Form 3 / Form 3B stereolithography printer with a nominal lateral and axial printing resolution of $25,\mu\mathrm{m}$, followed by standard post-processing (washing and curing) in accordance with the manufacturer’s recommendations. The individually printed components are shown in Fig.~\ref{fig:Fabrication}(a) and Fig.~\ref{fig:Fabrication}(d).

The next fabrication step involved forming the portion of the skull-conformal layer that attaches to the hologram surface. This interface was created using a progressive casting and UV-curing process\cite{flatlens}. Small volumes forming layers of BioMed Elastic 50A resin were added using a syringe and cured under 405 nm UV light, as shown in Fig.~\ref{fig:Fabrication}(b). This procedure was repeated until the complex features of the hologram were fully encased within the cured resin, as shown in Fig.~\ref{fig:Fabrication}(c). Between successive resin additions, hot air was applied to ensure that no air bubbles were introduced or trapped during curing. This step is essential for achieving homogeneity of the medium to ensure high acoustic performance of the lens. The next step in fabrication involved bonding the printed skull-conformal layer to the cast BioMed resin. This was achieved by applying a thin layer of BioMed Elastic 50A resin between the two components, followed by UV curing. The bottom surface, that bonds to the cast BioMed resin, of the printed skull-conformal layer (Fig.~\ref{fig:Fabrication}(d)) is designed to have slight convexity; this geometrical feature facilitates the outward flow of excess resin and entrapped air when pressure is applied. Under this applied pressure, the assembled system was placed in a UV curing chamber for $15,\mathrm{min}$.  All materials used in the fabrication are optically clear rather than opaque, allowing partial transmission of UV light through the lens and enabling effective curing of the thin bonding layer, thus obtaining the fabricated lens, as shown in Fig.~\ref{fig:Fabrication}(e). Finally, the fabricated lens was submerged in water and post-cured under UV light for 30 minutes at 70~$^\circ$C, as recommended by Formlabs\cite{formlabs_biomed_elastic_50a_guide}, to remove any uncured resin from the surface.

\subsection{Ex-vivo human skull specimen, preparation, and mounting}
Experiments were performed using ex-vivo human skull specimen (anatomical region: cranial bone; preservation state: formalin-fixed). Prior to acoustic measurements, each skull specimen was fully submerged in de-ionized water and degassed under vacuum to reduce trapped air within the porous cranial bone structure, which is known to introduce scattering and variability in trans-skull transmission. Degassing was performed at 28 mm of Hg (vacuum level) for $24\,\mathrm{h}$ at $28\,^\circ\mathrm{C}$, followed by equilibration in the temperature-controlled tank for $3\,\mathrm{h}$ before measurements. The tank setup with the holder was place in the tank for degassing at the same time. The skull was oriented such that the incident beam propagated through the region of interest along the planned trajectory.

A custom rigid holder was fabricated to secure the skull, ultrasound transducer, and the subject-specific holographic lens in a fixed relative geometry. The holder provided repeatable alignment features and prevented motion during scanning. The transducer position relative to the skull could be adjusted manually in $x$, $y$, and $z$ to set the desired standoff distance and alignment with the planned target. Registration between the physical experiment and the CT coordinate frame was achieved using holder based mechanical reference features, resulting in a repeatable mapping between the planned target and the tank/scan coordinates. Prior to each run, the lens and skull interfaces were inspected and gently agitated to remove entrapped air bubbles; the assembly was re-degassed for $3\,\mathrm{h}$ to ensure bubble-free coupling. The experimental setup is shown in Fig.~\ref{fig:experimental}(c).

\subsection{Pressure-field mapping}
\begin{figure*}[htp]
    \centering
    \includegraphics[width=1\linewidth]{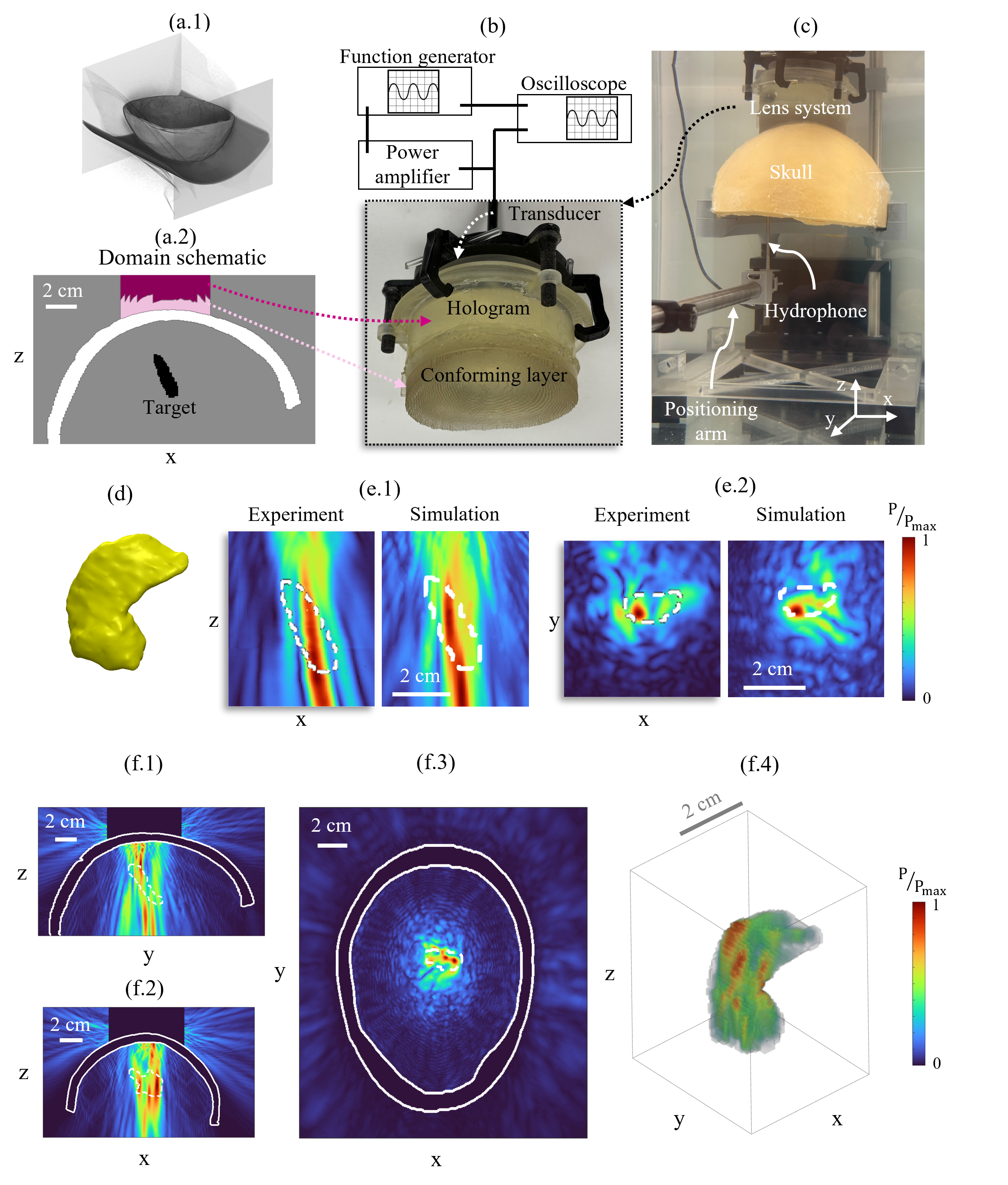}
    \caption{Workflow for subject-specific AHL design for experimental testing . (a.1) CT imaging for anatomical reconstruction. (a.2) Computational domain including holographic lens, conformal layer (CL), ex vivo skull, and target (hippocampus). (b) Schematic of the fabricated AHL system with transducer actuation. (c) Experimental setup with ex vivo skull mounted on a 3D positioning system. (d) 3D rendering of the Hippo. Comparison of experimental hydrophone scan and simulation results (e.1) in the sagittal plane, (e.2) in the transverse plane. (f) Full-wave time-domain computations of the acoustic pressure field, (f.1) coronal plane (yz), (f.2) sagittal plane (xz), and (f.3) transverse plane (xy). (f.4) volumetric pressure field distribution within the target.}
    \label{fig:experimental}
\end{figure*}

Ultrasound was generated using a custom single-element circular piezoelectric transducer with an active diameter of $50\,\mathrm{mm}$ and a nominal operating frequency of $675\,\mathrm{kHz}$. The transducer was fabricated in-house; details of the design and CAD model are shown in Fig.~\ref{fig:Fabrication}(f). The transducer was excited using a sinusoidal tone-burst waveform (10 cycles) produced by a Keysight 33500B signal generator and amplified by an Electronics \& Innovation A075 RF power amplifier. A peak-to-peak drive voltage of $0.7\,\mathrm{V_{pp}}$ was supplied to the transducer input. The lens-system axis was aligned to the planned trajectory by adjusting $x$--$y$--$z$ translation such that the acoustic axis passed through the CT-defined target. The lens system and the connections to the excitation instrumentation are shown in Fig.~\ref{fig:experimental}(b).

The transmitted acoustic field was measured on a specified target plane using an ONDA HNR--0500 needle hydrophone mounted on a custom-built in-house 3D positioning/scanning system, as shown in Fig.~\ref{fig:experimental}(c). Pressure-field maps were acquired over a scan area of $\,50\mathrm{mm}\times 50\mathrm{mm}$ with a spatial step size of $0.5\,\mathrm{mm}$ in the transverse plane and $\,50\mathrm{mm}\times 60\mathrm{mm}$ with a spatial step size of $0.5\,\mathrm{mm}$ in the sagittal plane. At each scan position a single burst of 10 cycles was recorded, and the oscilloscope acquisition window was set appropriately with a sampling rate of $1\,\mathrm{GS/s}$.

The hydrophone output was digitized using a Tektronix TBS2104 oscilloscope, and each waveform was averaged 64 times to improve signal-to-noise ratio. A MATLAB program was used to control the positioning system and synchronize scan motion with triggering of the function generator, ensuring consistent timing between excitation and sampling across the raster scan.

\subsection{Signal processing, waveform extraction, and pressure calibration.}
Recorded time-domain hydrophone signals were post-processed in MATLAB to remove DC offsets and suppress out-of-band electrical noise. For each scan position, the complex amplitude at the excitation frequency was extracted by computing the FFT of the time series and isolating the spectral component at $675\,\mathrm{kHz}$. The measured hydrophone voltage amplitude was converted to acoustic pressure using the manufacturer-provided calibration sensitivity for the ONDA HNR--0500 at the operating frequency.

\section{Experimental results}

Experimental analysis was conducted to validate the performance of the proposed system and assess its reliability under realistic operating conditions. Prior to performing the ex vivo propagation experiment, a full-wave simulation was carried out to validate the hologram lens and the HHO approach. Full-wave simulations were performed using the k-space pseudospectral method implemented in the open-source MATLAB toolbox k-Wave \cite{treeby2010k}. Although computationally demanding, this method has been demonstrated to be applicable in heterogeneous media for acoustic holography applications \cite{jimenez2019holograms, cengiz2024holographic, andres2022numerical}, and is therefore included in our SSVH framework. 

The optimal thickness map determined by the HHO method is used as input for the k-Wave simulations. Both the hologram and the secondary conforming layer are explicitly modeled. The head domain used in the MMDM simulations is also incorporated, and the acoustic field is forward-propagated to compute the resulting pressure distribution at the target volume. To achieve validation of the  of the full-wave simulations, pressure data are recorded only on three orthogonal planes centered at the target region. Pressure distributions recorded on the three orthogonal planes intersecting the target volume confirmed that the proposed design achieved effective phase compensation through the skull and soft tissue layers.

The computed thickness profile is embedded within the CAD design of the hologram housing, which facilitates coupling between the lens and the transducer. The objectives of the experiments were to (i) evaluate the accuracy and robustness of the HHO-optimized hologram design within the SSVH framework by comparing simulated and measured pressure fields, and (ii) demonstrate that a subject-specific acoustic holographic lens designed from CT-derived anatomy is clinically translatable.

The experimental measurements were limited to two-dimensional hydrophone scans acquired in the transverse and sagittal planes, and corresponding planes were extracted from the numerical simulation for direct comparison. The maximum experimental pressures measured in the transverse and sagittal planes were 122 kPa and 147 kPa, respectively. The transverse and sagittal planes acquired experimentally and the corresponding planes extracted from the numerical simulations are shown in Fig.~\ref{fig:experimental}(e.1) and (e.2). The agreement between the experimentally measured and simulated pressure fields was quantified using the structural similarity index (SSIM) and Mean absolute Error (MAE). The SSIM was computed using the built-in function available in MATLAB R2023a. The MAE is calculated as 
\begin{equation}
\mathrm{MAE}
=
\frac{1}{N}
\sum_{i=1}^{N}
\left|
P_{\mathrm{exp}}^{(i)} - P_{\mathrm{sim}}^{(i)}
\right|
\end{equation}

Further, the performance of the lens is quantified using focusing efficiency($\eta$) and area sonicated ($A_{\mathrm{son,t}}$). The focusing efficiency is calculated as:

\begin{equation}
\eta = 100 \times \frac{\Pi_{t}}{\Pi_{b}}=100\times \frac{\sum P_{t}^{2}}{\sum P_{b}^{2}}
\end{equation}
where $\eta$ denotes the focusing efficiency (\%) calculated at the experimental planes. The focusing efficiency indicates how effectively the acoustic energy is concentrated at the focal region, providing a measure of how well the system delivers energy to the intended target compared with off-target regions.

The area sonicated is calculates as:

\begin{equation}
A_{\text{son,t}}=
\sum_{i=1}^{n_x}
\sum_{j=1}^{n_y}
\left(
P_{ij} > P_{\text{th}}
\;\cdot\;
T_{ij} = 1
\right)
\cdot a
\end{equation}
where $T_{ij}$ is a binary mask defining the region that encompasses the target within the scanned plane, and $a$ is the unit pixel area, computed as $a = dx \cdot dy$. $A_{\mathrm{son,t}}$ represents the area sonicated within the target and is expressed in square centimeters (cm$^{2}$), while \(A_{\text{son,t}}(\%)=S_{\text{son,t}}/A_{t}\times100\) represents the percentage of the target region that is successfully sonicated, where $A_{t}$ denotes the total area of the target in the chosen plane.

 The MAE values were 0.052 and 0.053 for the transverse and sagittal planes, respectively, while the corresponding SSIM values were 0.40 and 0.56. Lower values of MAE indicate better agreement in pressure values, whereas higher SSIM values correspond to greater structural similarity between the fields. The performance metrics comparing experimental and simulated results are summarized in Table~\ref{tab:experiment_simulation}. The error metrics and sonication metrics indicate a good agreement between the experimental and simulated pressure fields in both planes, with consistent spatial correspondence of the focal region and the surrounding pressure distribution.  This agreement demonstrates that Hounsfield units based thresholding of the CT scan provided a sufficient geometric accuracy for the HHO method to account for phase aberrations. Furthermore, results validate the end-to-end workflow from imaging, optimization, fabrication, to acoustic verification. Furthermore, the employed fabrication processes were found to perform as intended, yielding experimental performance in close agreement with the numerical model, thereby supporting the clinical translatability of the fabrication workflow.

\begin{table}[ht]
\centering
\caption{Summary of the targeting results from experimental scan compared with full-wave simulations. SSIM = structural similarity index measure, MAE = Mean Absolute Error, $\eta$ = focusing efficiency, $A_{\mathrm{son,t}}(cm^2)$ = sonicated area $(>115~\mathrm{kPa})$, $A_{\mathrm{son,t}}$ = \% of the target successfully sonicated}
\label{tab:experiment_simulation}
\footnotesize
\setlength{\tabcolsep}{5pt}
\renewcommand{\arraystretch}{1.1}

\begin{tabular}{llcc}
\hline
\toprule
\textbf{Anatomical Plane} & \textbf{Metric} & \textbf{Experiment} & \textbf{Simulation} \\
\midrule
\hline

\multirow{4}{*}{Transversal}
& SSIM  
& \multicolumn{2}{c}{0.40} \\
& MAE  
& \multicolumn{2}{c}{0.052} \\
& $\eta$ (\%) 
& 47.7 & 50 \\
& $A_{\mathrm{son,t}}$ $(\mathrm{cm}^2)$
& 0.95 & 0.94 \\
& $A_{\mathrm{son,t}}$ (\%) 
& 42.6 & 45.3 \\
\addlinespace
\hline
\multirow{4}{*}{Sagittal}
& SSIM 
& \multicolumn{2}{c}{0.56} \\
& MAE 
& \multicolumn{2}{c}{0.053} \\
& $\eta$ (\%) 
& 27.7 & 30.6 \\
& $A_{\mathrm{son,t}}$ $(\mathrm{cm}^2)$
& 1.94 & 1.67 \\
& $A_{\mathrm{son,t}}$ (\%) 
& 62.8 & 57.9 \\
\bottomrule
\hline
\end{tabular}
\end{table}

 Given the strong correspondence observed in these orthogonal two-dimensional sections, it can be reasonably inferred that the volumetric sonication metrics computed from the three-dimensional simulation are representative of the experimental behavior. As direct experimental measurement of the full three-dimensional pressure field was not available due to scanning limitations, the volumetric metrics derived from the numerical model are therefore reported as a close approximation to the experimental sonication characteristics. For the deep hippocampal target, the simulation predicted a sonicated volume within the target $V_{\mathrm{son,t}}$ of 4.6~cm$^{3}$, corresponding to 51.25\% of the target volume. The target volume fraction above the pressure threshold ($V_{\mathrm{tvf}}$) was 25.2\%, and the associated amplitude demand ratio (ADR) was 268.33. The simulated pressure fields shown correspond to centroidal planes extracted from the three-dimensional domain, as illustrated in Fig.~\ref{fig:experimental}(f). The differences between the experimental and numerical results could be attributed to factors such as alignment challenges and variations in skull properties, mainly the density and the speed of sound due to the porous structure of the skull. The vacuum chamber process, explained in Section IV.B, reduced the presence of air within the porous structure of the skull, causing it to behave more closely to the properties used in the numerical modeling.



\section{Conclusion}

This study establishes a subject-specific volumetric holography (SSVH) framework for acoustic holography–based ultrasound delivery in transcranial focused ultrasound (tFUS) neuromodulation. Central to this approach is a skull- and scalp-conforming acoustic holographic architecture, coupled with a subject-specific fabrication workflow, which physically integrates individualized phase correction with impedance-matched anatomical conformity. Combined with a physics-guided holographic optimization methodology that embeds individual cranial geometry and therapy-driven constraints, this design enables precise three-dimensional wavefield synthesis through heterogeneous skull structures.

Computational analyses across 20 cases, spanning 5 subjects and 4 brain regions at varying depths—including deep, mid-depth, and superficial targets—demonstrate robust and reproducible volumetric focusing. On average, 45\% $\pm$ 9.6 of the prescribed target volume achieved pressure amplitudes at or above 115~kPa, with a targeted volume fraction of 23\% $\pm$ 2.5, while maintaining exposure within established safety limits. These results confirm effective spatial confinement and reliable targeting despite skull-induced scattering and phase aberrations.

Experimental validation using hydrophone measurements through an ex vivo human skull confirms accurate reconstruction of the synthesized pressure fields, with a mean absolute error of 0.052, demonstrating strong agreement between simulation and experiment.

By embedding wavefront engineering directly into an anatomically conformal interface, this work establishes a scalable and mechanically robust platform that unites physics-driven optimization with boundary-condition control. This framework advances tFUS from conventional electronic beam steering toward physically encoded, geometry-aware acoustic modulation, opening a pathway for compact, high-fidelity, and clinically deployable ultrasound systems capable of precision neuromodulation and broader therapeutic acoustic interventions. 

\begin{acknowledgments}
C.C. and M.P. contributed equally and shared first-authorship. 
This work was supported by the U.S. National Science
Foundation (NSF) under the grant, Award No.
CAREER CMMI 2143788, which is gratefully acknowledged.
\end{acknowledgments}

\noindent\textbf{Conflict of Interest
}

The authors declare no conflict of interest.

\noindent\textbf{Data Availability statement
}

The data that support the findings of this study are available from the corresponding author upon reasonable request.

\newpage

\bibliography{ref}
\end{document}